# Estimating the Healthy Life Expectancy from the Health State Function of a Population in Connection to the Life Expectancy at Birth

## Christos H Skiadas and Charilaos Skiadas

Technical University of Crete, Chania, Crete, Greece
E-mail: skiadas@cmsim.net
Hanover College, Indiana, USA
E-mail: skiadas@hanover.edu

**Abstract:** Following our previous works on the health state of a population and the related health state function we proceed in developing a method to estimate the Healthy Life Expectancy in connection to the relative impact of the Mortality Area in the health state function graph.
The resulting tools are applied to the data sets for 1990, 2000 and 2009 for the Countries of the World Health Organization (WHO). The application is done in the Excel Chart and it is ready to be used for other time periods.
The results are compared with the estimates presented in the WHO report for 2000 showing a good relationship between the estimates of the two methods. However, our proposed method, not based on collection of data for diseases and other causes affecting a healthy life, is more flexible. We can estimate the healthy life years from various time periods when information related to diseases is missing.

## Introduction

In previous studies (Skiadas Oct. 2011, Dec. 2011, Feb. 2012 and Skiadas & Skiadas 2007, 2010 a, b, 2011) we have introduced the Health State Function (HSF) of a Population and we have applied to population and mortality data in various countries. In these studies we have further analyzed and expanded the theory proposed by Janssen and Skiadas (1995). Following the traditional methods (Graunt 1676, Halley 1693 and Gompertz 1825) developed in Demography and Actuarial Science we have proposed methods and techniques to estimate the HSF directly from the data sets, without fitting any intermediate function as is also the case for finding the mortality $\mu_x$ from the same data sets. We have also suggested a method for presenting both HSF and $\mu_x$ in the same graph (Skiadas 2012).

---

Having introduced a function expressing the development of the health state of a population according to the age of this population we can estimate several interesting and important characteristics related to the health state. The graph for the HSF versus age (see Figure 1) is of a non-symmetric parabola form showing a plateau like shape during 30-45 years and then slowly decreasing until a fast decrease stage at old ages. (In our notation we use the term $t$ for the age instead of $x$ preferred by actuaries. This is because the same theory and practice can apply to other disciplines and especially when estimating the life time of machines and complex equipments).

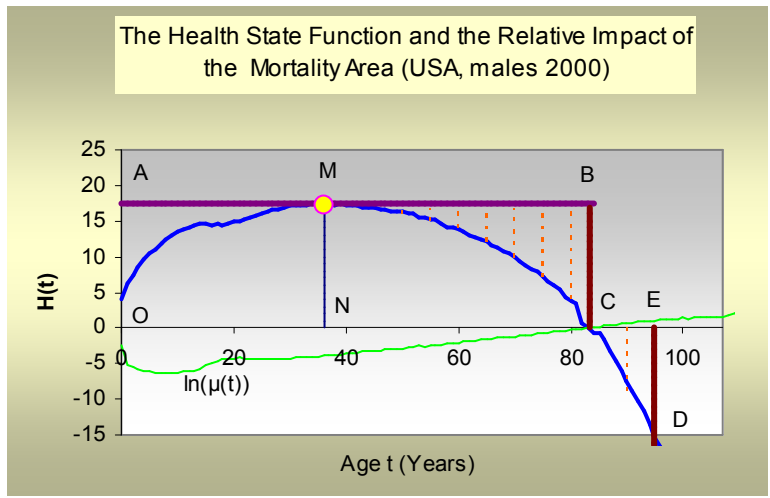

Fig. 1. The impact of the mortality area to health state

By observing the above graph we can immediately see that the area between the health state curve and the horizontal axis (OMCO) represents the total health dynamics (THD) of the population. Of particular importance is also the area of the health rectangle (OABC) which includes the health state curve. This rectangle is divided in two rectangular parts the smaller (OAMN) indicating the first part of the human life until reaching the point M at the highest level of health state (usually the maximum is between 30 to 45 years) and the second part (NMBC) characterized by the gradual deterioration of the human organism until the zero level of the health state. This zero point health age C is associated with the maximum death rate. After this point the health state level appears as negative in the graph and characterizes a part of the human life totally unstable with high mortality; this is also indicated by a positively increasing form of the logarithm of the force of mortality $\ln(\mu_x)$.

We call the second rectangle NMBC as the ***deterioration rectangle***. Instead the first rectangle OAMN is here called as the ***development rectangle***. For both cases we can find the relative impact of the area inside each rectangle but outside the health state area to the overall health state. In this paper we analyze the relative impact of the ***deterioration area*** MBCM indicated by dashed lines in the ***deterioration rectangle***. It should be noted that if no-deterioration mechanism was present or the repairing mechanism was perfect the health state should continue following the straight line AMB parallel to the X-axis at the level of the maximum health state. The smaller the deterioration area related to the health state area, the higher the healthy life of the population. This comparison can be done by estimating the related areas and making a simple division.

However, when trying to expand the human life further than the limits set by the deterioration mechanisms the percentage of the non-healthy life years becomes higher. This means that we need to divide the total rectangle area by that of the deterioration area to find an estimate for the "lost healthy life years". It is clear that if we don't correct the deterioration mechanisms the loss of healthy years will become higher as the expectation of life becomes larger. This is already observed in the estimates of the World Health Organization (WHO) in the World Health Report for 2000 where the lost healthy years for females are higher than the corresponding values for males. The females show higher life expectancy than males but also higher values for the lost healthy years. The proposed "loss of healthy life years" indicator is given by:

$$LHLY_1 = \lambda \frac{OABC}{THD_{ideal}} \cdot \frac{THD_{ideal}}{MBCM} = \lambda \frac{OABC}{MBCM}$$

Where $THD_{ideal}$ is ideal total health dynamics of the population and the parameter $\lambda$ expresses years and should be estimated according to the specific case. For comparing the related results in various countries we can set $\lambda=1$. When OABC approaches the $THD_{ideal}$ as is the case of several countries in nowadays the loss of healthy life years indicator LHLY can be expressed by other forms.

Another point is the use of the (ECD) area in improving forecasts especially when using the 5-year life tables as is the case of the data for all the WHO Countries. In this case the expanded loss of healthy life years indicator LHLY will take the following two forms:

$$LHLY_2 = \lambda \frac{OMCO + ECD}{MBCM}$$

$$LHLY_3 = \lambda \frac{OABC + ECD}{MBCM}$$

It is clear that the last form will give higher values than the previous one. The following scheme applies: $LHLY_1 < LHLY_2 < LHLY_3$. It remains to explore the forecasting ability of the three forms of the "loss of healthy life years" indicator by applying LHLY to life tables provided by WHO or by the Human Mortality Database or by other sources.

**Related Theory**

Life expectancy and life expectancy at birth are indicators that should be related to the health state of a population. However, a crucial aspect in defining the connection between the health state and life expectancy of a population was the luck of a quantitative definition and measure of the health state of a population. Recently we have succeeded in estimating the health state of a population by using the data sets for the yearly deaths per year of age and the age distribution of the population for the same year. The used data are provided by the human mortality database for 35 countries.

The indicator for the health state of a population is based (Skiadas 2012) on the following probability density function $g(t)$

$$g(t) = \frac{k}{\sqrt{t^3}} e^{-\frac{(H_t)^2}{2t}}$$

Where $H_t$ is the Health State Function and $k$ is a parameter and $g(t)$ is given by using the data sets provided by the data bases. A simple estimate is achieved by an Excel program (SKI-6-Parameters) provided in the website http://www.cmsim.net .

The Health State is found by solving the above relation for $H_t$. The resulting form is:

$$H_t = \left| \left( -2t \ln \frac{g(t)\sqrt{t^3}}{k} \right)^{1/2} \right|$$

The parameter $k$ is defined by the relation

$$k = \max \left( g(t)\sqrt{t^3} \right)$$

The total estimate of the health state of the population is given by the relation

$$H_{total} = \int_0^{t(H=0)} H_t dt \approx \sum_0^{t(H=0)} H_t \Delta_t$$

Where $t(H=0)$ is the age of the maximum death rate corresponding to zero health state $H_t=0$ and $\Delta_t=(t+j-t)$ where $j$ is the interval between age groups. The total health state $H_{total}$ accounts for the surface covered by the horizontal axis $X$ and the positive part of the health state function (see Figure 1). It must be noted that the negative part of the health state function do not produce stochastic paths for the health state as is already known from the theory of the first exit time of a stochastic process. This is clearly illustrated in the next Figure 2 for males in USA (2000).

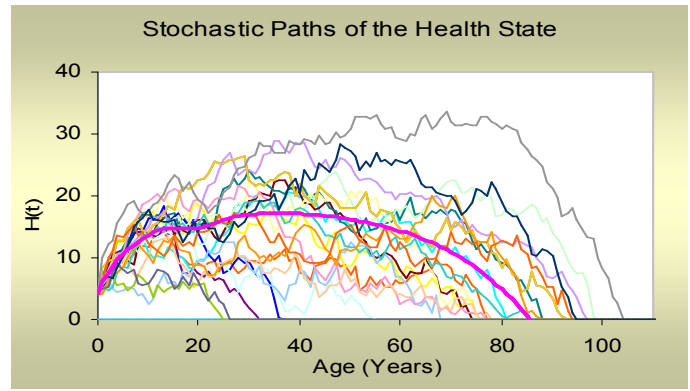

Fig. 2. Health state function and stochastic paths

Other indicators are the age of the maximum health state and the maximum health state. Both estimates are not very precise due to the almost flat shape of the health state curve around the maximum (see Figure 1). Instead the total health state $H_{total}$ accounts for the main time period of the healthy life time expressed by the positive part of the health state function.

The ranking of 36 Countries (males and females) selected from the Human Mortality Database according to their scores for the total health state $H_{total}$ (THS) the age 2000 are given in the next Table I. The related values for the Life Expectancy at Birth (LEB=$e_x$) are also included in the same Table along with the estimated LEB after finding the relationship between THS and LEB which follows a linear function for males (see Figure 3) and females (see Figure 4) of the form LEB=34.88+0.0352*THS and LEB=39.67+0.0307*THS respectively. As was expected the total health state for females is higher than males in the same country and at the same year.

The next step is to find the Life Expectancy at Birth versus the Total Health State during time for the same country. The cases for males and females in USA from 1933 to 2007 are explored. That it is found is a linear relationship for the first period from 1933 to 1960 for males (slope 0.0785) and another linear part from 1960 to 2007 with lower slope (0.0364). For females the slope is 0.0517 for the first period and 0.0394 for the second period. ***The findings are in favor of a slower development of the life expectancy at birth related to the growth of the total health state of the population***. The argument is further explored by using the data for Sweden (females) from 1751 to 2007. A sigmoid like function is applied indicating that the life expectancy at birth merely tends to a level even if the total health state tends to increase considerably during time (see Figure 5).

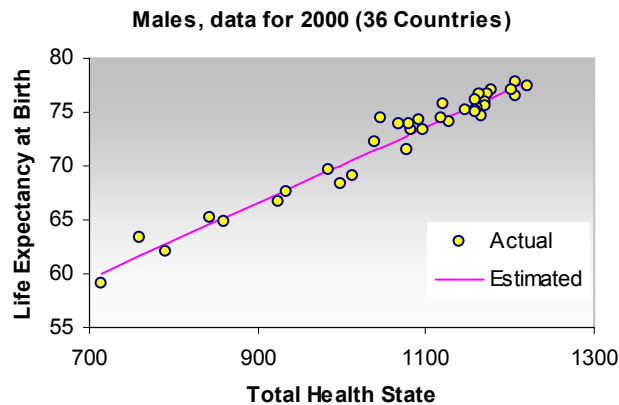

Fig. 3. Life expectancy at birth versus total health state

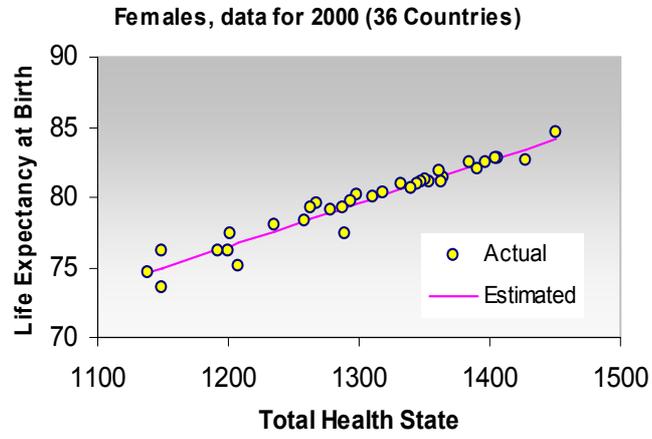

Fig. 4. Life expectancy at birth versus total health state

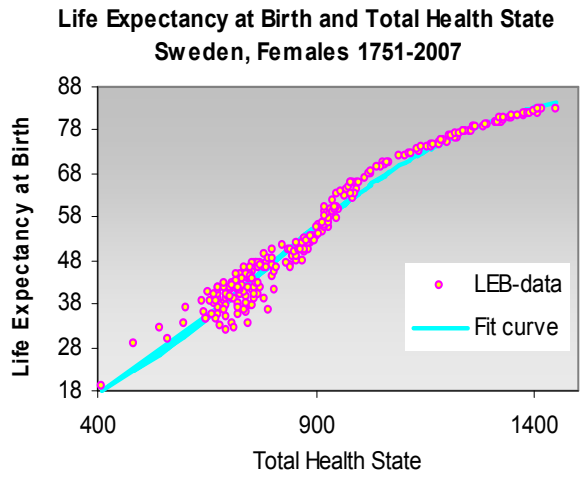

Fig. 5. Life expectancy at birth versus total health state in Sweden

8     C H Skiadas and C Skiadas

| TABLE I | | | | | | | |
|---|---|---|---|---|---|---|---|
| Total Health State (THS) and Life Expectancy at Birth (LEB) for various Countries | | | | | | | |
| Males | THS | LEB | LEB Estimated | Females | THS | LEB | LEB Estimated |
| Sweden | 1220 | 77,40 | 77,85 | Japan | 1451 | 84,58 | 84,18 |
| Italy | 1207 | 76,55 | 77,39 | Switzerland | 1427 | 82,62 | 83,45 |
| Japan | 1206 | 77,70 | 77,36 | Spain | 1406 | 82,72 | 82,80 |
| Australia | 1202 | 77,12 | 77,22 | France | 1405 | 82,80 | 82,77 |
| Switzerland | 1178 | 76,95 | 76,37 | Italy | 1397 | 82,50 | 82,53 |
| Canada | 1174 | 76,64 | 76,23 | Sweden | 1391 | 82,04 | 82,34 |
| Norway | 1172 | 75,95 | 76,16 | Australia | 1385 | 82,41 | 82,16 |
| Netherlands | 1170 | 75,53 | 76,09 | Norway | 1365 | 81,37 | 81,54 |
| Belgium | 1165 | 74,58 | 75,91 | Austria | 1363 | 81,10 | 81,48 |
| Israel | 1163 | 76,69 | 75,84 | Canada | 1361 | 81,83 | 81,42 |
| UK | 1161 | 75,43 | 75,77 | Germany | 1354 | 81,01 | 81,21 |
| New Zealand | 1159 | 76,09 | 75,70 | New Zealand | 1350 | 81,24 | 81,08 |
| France | 1158 | 75,24 | 75,67 | Finland | 1347 | 81,02 | 80,99 |
| Germany | 1158 | 74,96 | 75,67 | Belgium | 1345 | 80,91 | 80,93 |
| Austria | 1146 | 75,10 | 75,25 | Netherlands | 1340 | 80,57 | 80,78 |
| Finland | 1129 | 74,15 | 74,65 | Israel | 1333 | 81,00 | 80,56 |
| Spain | 1120 | 75,77 | 74,33 | Portugal | 1318 | 80,30 | 80,10 |
| Denmark | 1119 | 74,44 | 74,29 | Chile | 1311 | 79,96 | 79,89 |
| Portugal | 1098 | 73,30 | 73,55 | UK | 1298 | 80,23 | 79,49 |
| USA | 1093 | 74,19 | 73,38 | Slovenia | 1294 | 79,76 | 79,37 |
| Taiwan | 1083 | 73,32 | 73,03 | Slovakia | 1289 | 77,33 | 79,21 |
| Ireland | 1081 | 73,98 | 72,96 | Ireland | 1287 | 79,23 | 79,15 |
| Czech Republic | 1078 | 71,56 | 72,85 | Denmark | 1279 | 79,12 | 78,91 |
| Chile | 1067 | 73,89 | 72,46 | USA | 1268 | 79,52 | 78,57 |
| Luxemburg | 1047 | 74,45 | 71,76 | Taiwan | 1263 | 79,22 | 78,42 |
| Slovenia | 1039 | 72,18 | 71,48 | Czech Republic | 1258 | 78,34 | 78,26 |
| Slovakia | 1013 | 69,02 | 70,56 | Poland | 1236 | 77,95 | 77,59 |
| Bulgaria | 998 | 68,30 | 70,03 | Bulgaria | 1208 | 75,01 | 76,73 |
| Poland | 985 | 69,57 | 69,57 | Lithuania | 1201 | 77,37 | 76,51 |
| Hungary | 934 | 67,53 | 67,78 | Latvia | 1200 | 76,08 | 76,48 |
| Lithuania | 925 | 66,74 | 67,46 | Hungary | 1193 | 76,08 | 76,27 |
| Latvia | 860 | 64,87 | 65,17 | Estonia | 1149 | 76,20 | 74,92 |
| Estonia | 843 | 65,20 | 64,57 | Ukraine | 1149 | 73,52 | 74,92 |
| Ukraine | 791 | 62,08 | 62,74 | Belarus | 1138 | 74,67 | 74,58 |
| Belarus | 759 | 63,31 | 61,61 | Luxemburg | 1129 | 80,74 | 74,31 |
| Russia | 714 | 58,99 | 60,03 | Russia | 1080 | 72,25 | 72,80 |

**Application to WHO data**

A complete data set for almost all the countries of the World can be found and download from the WHO database in Excel format. For our application here we use the LT199020002009whs2011 package downloaded from the WHO database and including the Life Tables for all WHO Countries for 1990, 2000 and 2009.

We expand the Table (see Table II) to the right by estimating the "Loss of Healthy Life Years" ($LHLY_1$, $LHLY_2$ and $LHLY_3$) in the columns next to the Life Expectancy column (see Table II). After estimating the LHLY we estimate the Healthy Life Years ($HALE_1$, $HALE_2$ and $HALE_3$) from $HALE = e_x - LHLY$, where $e_x$ is the Life Expectancy. In the same columns we estimate the HALE for all the age period of the human life as is the case for $e_x$. We use the data $_nd_x$ form the column I as starting values of the calculations beginning from column S.

Other very important indicators are also estimated in the related columns. The values for $k(t)$ are provided in column W and the maximum $k_{max}$ is given in column X. The Health State Function values are given in column AB whereas the maximum Health State is given in column AC. The Expected Healthy Age is given in column AO. The expected healthy age is estimated by

$$\overline{T} = \int_0^{t(H=0)} tH_t^* dt \approx \sum_0^{t(H=0)} tH_t^* \Delta_t$$

Where $H^*$ is the normalized value of the Health State calculated in column AN and $t(H=0)$ is the age where the Health State is zero. Another characteristic indicator is given in column AP by multiplying the maximum health state with the expected health age. The resulting indicator characterizes the health status of the population. Finally a very important indicator is given in column AR. This indicator is measuring the Total Health State of the population and it is expressed by the area (OMCO) in Figure 1. It is estimated by the following formula:

$$H_{total} = \int_0^{t(H=0)} H_t dt \approx \sum_0^{t(H=0)} H_t \Delta_t$$

The Total Health State indicator provides another tool to classify the various countries. The rankings estimated are analogues to the life expectancy at birth. However, this indicator can give more precise results. Even more, as is the case when introducing new methods and tools, new frontiers open for analysis and research.

Table II

| A | B | C | D | E | F | G | H | I | J | K | L | M | N | O | P |
|---|---|---|---|---|---|---|---|---|---|---|---|---|---|---|---|
| iso | country | year | sex | age $x$ | $_nM_x$ | $_nq_x$ | $l_x$ | $_nd_x$ | $_nL_x$ | $T_x$ | $e_x$ | LHLY3 | HALE3 | LHLY2 | HALE2 |
| AFG | Afghanistan | 2009 | males | 0 | 0,15973 | 0,14366 | 100000 | 14.366 | 89.944 | 4.656.098 | 46,6 | 11,3 | 35,3 | 9,7 | 36,9 |
| AFG | Afghanistan | 2009 | males | 1 | 0,01980 | 0,07561 | 85634 | 6.475 | 326.995 | 4.566.154 | 53,3 |  | 40,4 |  | 42,2 |
| AFG | Afghanistan | 2009 | males | 5 | 0,00438 | 0,02166 | 79159 | 1.714 | 391.509 | 4.239.159 | 53,6 |  | 40,6 |  | 42,4 |
| AFG | Afghanistan | 2009 | males | 10 | 0,00227 | 0,01127 | 77445 | 873 | 385.042 | 3.847.650 | 49,7 |  | 37,6 |  | 39,3 |
| AFG | Afghanistan | 2009 | males | 15 | 0,00183 | 0,00909 | 76572 | 696 | 381.122 | 3.462.607 | 45,2 |  | 34,2 |  | 35,8 |
| AFG | Afghanistan | 2009 | males | 20 | 0,00562 | 0,02769 | 75876 | 2.101 | 374.130 | 3.081.486 | 40,6 |  | 30,8 |  | 32,1 |
| AFG | Afghanistan | 2009 | males | 25 | 0,00758 | 0,03720 | 73775 | 2.744 | 362.017 | 2.707.356 | 36,7 |  | 27,8 |  | 29,0 |
| AFG | Afghanistan | 2009 | males | 30 | 0,00924 | 0,04515 | 71031 | 3.207 | 347.138 | 2.345.340 | 33,0 |  | 25,0 |  | 26,1 |
| AFG | Afghanistan | 2009 | males | 35 | 0,01123 | 0,05462 | 67824 | 3.704 | 329.860 | 1.998.201 | 29,5 |  | 22,3 |  | 23,3 |
| AFG | Afghanistan | 2009 | males | 40 | 0,01334 | 0,06457 | 64120 | 4.140 | 310.249 | 1.668.342 | 26,0 |  | 19,7 |  | 20,6 |
| AFG | Afghanistan | 2009 | males | 45 | 0,01635 | 0,07853 | 59980 | 4.710 | 288.123 | 1.358.093 | 22,6 |  | 17,1 |  | 17,9 |
| AFG | Afghanistan | 2009 | males | 50 | 0,02098 | 0,09969 | 55269 | 5.510 | 262.573 | 1.069.970 | 19,4 |  | 14,7 |  | 15,4 |
| AFG | Afghanistan | 2009 | males | 55 | 0,02976 | 0,13850 | 49760 | 6.892 | 231.569 | 807.397 | 16,2 |  | 12,3 |  | 12,8 |
| AFG | Afghanistan | 2009 | males | 60 | 0,03929 | 0,17886 | 42868 | 7.668 | 195.171 | 575.828 | 13,4 |  | 10,1 |  | 10,6 |
| AFG | Afghanistan | 2009 | males | 65 | 0,05483 | 0,24110 | 35200 | 8.487 | 154.785 | 380.657 | 10,8 |  | 8,2 |  | 8,5 |
| AFG | Afghanistan | 2009 | males | 70 | 0,08127 | 0,33774 | 26714 | 9.022 | 111.012 | 225.872 | 8,5 |  | 6,4 |  | 6,7 |
| AFG | Afghanistan | 2009 | males | 75 | 0,11943 | 0,45985 | 17691 | 8.135 | 68.118 | 114.860 | 6,5 |  | 4,9 |  | 5,1 |
| AFG | Afghanistan | 2009 | males | 80 | 0,17730 | 0,61423 | 9556 | 5.870 | 33.106 | 46.742 | 4,9 |  | 3,7 |  | 3,9 |
| AFG | Afghanistan | 2009 | males | 85 | 0,25197 | 0,77295 | 3686 | 2.849 | 11.308 | 13.636 | 3,7 |  | 2,8 |  | 2,9 |
| AFG | Afghanistan | 2009 | males | 90 | 0,34413 | 0,84661 | 837 | 709 | 2.059 | 2.328 | 2,8 |  | 2,1 |  | 2,2 |
| AFG | Afghanistan | 2009 | males | 95 | 0,46466 | 0,88463 | 128 | 114 | 244 | 269 | 2,1 |  | 1,6 |  | 1,7 |
| AFG | Afghanistan | 2009 | males | 100 | 0,60981 | 1 | 15 | 15 | 24 | 24 | 1,6 |  | 1,2 |  | 1,3 |
| AFG | Afghanistan | 2009 | females | 0 | 0,13477 | 0,12316 | 100000 | 12.316 | 91.379 | 5.024.534 | 50,2 | 12,3 | 37,9 | 10,5 | 39,7 |
| AFG | Afghanistan | 2009 | females | 1 | 0,01955 | 0,07470 | 87684 | 6.550 | 335.017 | 4.933.155 | 56,3 |  | 42,5 |  | 44,6 |
| AFG | Afghanistan | 2009 | females | 5 | 0,00447 | 0,02210 | 81134 | 1.793 | 401.188 | 4.598.138 | 56,7 |  | 42,8 |  | 44,9 |
| AFG | Afghanistan | 2009 | females | 10 | 0,00267 | 0,01327 | 79341 | 1.053 | 394.072 | 4.196.950 | 52,9 |  | 39,9 |  | 41,9 |
| AFG | Afghanistan | 2009 | females | 15 | 0,00385 | 0,01907 | 78288 | 1.493 | 387.706 | 3.802.878 | 48,6 |  | 36,7 |  | 38,5 |
| AFG | Afghanistan | 2009 | females | 20 | 0,00560 | 0,02761 | 76795 | 2.120 | 378.673 | 3.415.171 | 44,5 |  | 33,6 |  | 35,2 |
| AFG | Afghanistan | 2009 | females | 25 | 0,00628 | 0,03093 | 74675 | 2.310 | 367.598 | 3.036.498 | 40,7 |  | 30,7 |  | 32,2 |

Table II (continued…)

| Q | R | S | T | U | V | W | X | Y | Z | AA | AB | AC | AD | AE | AF |
|---|---|---|---|---|---|---|---|---|---|---|---|---|---|---|---|
| LHLY | HALE1 | dx/Sum(dx) | dx-normalized | | | kx (max) | | | | | Health State | Max Health | | | |
| 7,2 | 39,4 | 0,1437 | 0,1437 | 1 | 0,14366 | 0,1437 | 10,79 | 0,1437 | 2,939 | 0 | 2,93917799 | 12,39 | 2,9392 | 0 | 0 |
| | 45,1 | 0,0647 | 0,0647 | 4 | 0,01619 | 0,0458 | | 0,1831 | 4,675 | 0 | 4,67455791 | | 4,6746 | 0 | 0 |
| | 45,4 | 0,0171 | 0,0171 | 5 | 0,00343 | 0,0504 | | 0,2519 | 8,025 | 0 | 8,02537239 | | 8,0254 | 0 | 0 |
| | 42,1 | 0,0087 | 0,0087 | 5 | 0,00175 | 0,0637 | | 0,3185 | 10,626 | 0 | 10,6263154 | | 10,6263 | 0 | 0 |
| | 38,3 | 0,0070 | 0,0070 | 5 | 0,00139 | 0,0891 | | 0,4454 | 12,390 | 0 | 12,3899422 | | 12,3899 | 15 | 15 |
| | 34,4 | 0,0210 | 0,0210 | 5 | 0,00420 | 0,4044 | | 2,0219 | 11,745 | 0 | 11,7451492 | | 11,7451 | 0 | 15 |
| | 31,1 | 0,0274 | 0,0274 | 5 | 0,00549 | 0,7276 | | 3,6378 | 11,843 | 0 | 11,8427389 | | 11,8427 | 0 | 15 |
| | 27,9 | 0,0321 | 0,0321 | 5 | 0,00641 | 1,1070 | | 5,5352 | 11,883 | 0 | 11,8826296 | | 11,8826 | 0 | 15 |
| | 25,0 | 0,0370 | 0,0370 | 5 | 0,00741 | 1,6001 | | 8,0006 | 11,724 | 0 | 11,7238039 | | 11,7238 | 0 | 15 |
| | 22,0 | 0,0414 | 0,0414 | 5 | 0,00828 | 2,1737 | | 10,8686 | 11,464 | 0 | 11,463674 | | 11,4637 | 0 | 15 |
| | 19,1 | 0,0471 | 0,0471 | 5 | 0,00942 | 2,9389 | | 14,6944 | 10,941 | 0 | 10,9405346 | | 10,9405 | 0 | 15 |
| | 16,4 | 0,0551 | 0,0551 | 5 | 0,01102 | 4,0136 | | 20,0679 | 10,046 | 0 | 10,0457475 | | 10,0457 | 0 | 15 |
| | 13,7 | 0,0689 | 0,0689 | 5 | 0,01378 | 5,7763 | | 28,8817 | 8,369 | 0 | 8,36858793 | | 8,3686 | 0 | 15 |
| | 11,3 | 0,0767 | 0,0767 | 5 | 0,01534 | 7,3064 | | 36,5319 | 6,901 | 0 | 6,90064826 | | 6,9006 | 0 | 15 |
| | 9,1 | 0,0849 | 0,0849 | 5 | 0,01697 | 9,1011 | | 45,5057 | 4,746 | 0 | 4,7463878 | | 4,7464 | 0 | 15 |
| | 7,2 | 0,0902 | 0,0902 | 5 | 0,01804 | 10,7948 | | 53,9742 | 0,000 | 70 | 0 | | 0,0000 | 0 | 15 |
| | 5,5 | 0,0813 | 0,0813 | 5 | 0,01627 | 10,7796 | | 53,8981 | 0,463 | 0 | -0,46296406 | | 0,0000 | 0 | 15 |
| | 4,1 | 0,0587 | 0,0587 | 5 | 0,01174 | 8,5584 | | 42,7919 | 6,133 | 0 | -6,13265955 | | 0,0000 | 0 | 15 |
| | 3,1 | 0,0285 | 0,0285 | 5 | 0,00570 | 4,5443 | | 22,7214 | 12,199 | 0 | -12,1989286 | | 0,0000 | 0 | 15 |
| | 2,4 | 0,0071 | 0,0071 | 5 | 0,00142 | 1,2309 | | 6,1547 | 19,879 | 0 | -19,8790326 | | 0,0000 | 0 | 15 |
| | 1,8 | 0,0011 | 0,0011 | 5 | 0,00023 | 0,2145 | | 1,0723 | 27,430 | 0 | -27,4298071 | | 0,0000 | 0 | 15 |
| | 1,4 | 0,0001 | 0,0001 | 5 | 0,00003 | 0,0305 | | 0,1523 | 34,437 | 0 | -34,4366585 | | 0,0000 | 0 | 15 |
| 8,6 | 41,6 | 0,1232 | 0,1232 | 1 | 0,12316 | 0,1232 | 13,57 | 0,1232 | 3,067 | 0 | 3,06661523 | 13,30 | 3,0666 | 0 | 0 |
| | 46,7 | 0,0655 | 0,0655 | 4 | 0,01638 | 0,0463 | | 0,1853 | 4,767 | 0 | 4,76658043 | | 4,7666 | 0 | 0 |
| | 47,0 | 0,0179 | 0,0179 | 5 | 0,00359 | 0,0527 | | 0,2635 | 8,162 | 0 | 8,16152356 | | 8,1615 | 0 | 0 |
| | 43,9 | 0,0105 | 0,0105 | 5 | 0,00211 | 0,0768 | | 0,3842 | 10,669 | 0 | 10,6689334 | | 10,6689 | 0 | 0 |
| | 40,3 | 0,0149 | 0,0149 | 5 | 0,00299 | 0,1911 | | 0,9555 | 11,679 | 0 | 11,6793576 | | 11,6794 | 0 | 0 |
| | 36,9 | 0,0212 | 0,0212 | 5 | 0,00424 | 0,4080 | | 2,0402 | 12,132 | 0 | 12,1316313 | | 12,1316 | 0 | 0 |
| | 33,7 | 0,0231 | 0,0231 | 5 | 0,00462 | 0,6125 | | 3,0625 | 12,692 | 0 | 12,692379 | | 12,6924 | 0 | 0 |

Table II (continued)

| AG | AH | AI | AJ | AK | AL | AM | AN | AO | AP | AQ | AR |
|---|---|---|---|---|---|---|---|---|---|---|---|
| | | | | | | | | Exp Health | lax*Exp | Health State | Total |
| 0 | 0,000 | 0,000 | 2,939 | 2,939 | 0,000 | 0,0212 | 0,0212 | 32,86 | 407 | 2,93918 | 675,1 |
| 0 | 0,000 | 0,000 | 18,698 | 18,698 | 0,000 | 0,0338 | 0,1014 | | | 18,69823 | |
| 0 | 0,000 | 0,000 | 40,127 | 40,127 | 0,000 | 0,0580 | 0,4062 | | | 40,12686 | |
| 0 | 0,000 | 0,000 | 53,132 | 53,132 | 0,000 | 0,0768 | 0,9219 | | | 53,13158 | |
| 15 | 0,000 | 61,950 | 61,950 | 61,950 | 0,000 | 0,0896 | 1,5228 | | | 61,94971 | |
| 15 | 3,224 | 58,726 | 58,726 | 58,726 | 0,000 | 0,0849 | 1,8681 | | | 58,72575 | |
| 15 | 2,736 | 59,214 | 59,214 | 59,214 | 0,000 | 0,0856 | 2,3118 | | | 59,21369 | |
| 15 | 2,537 | 59,413 | 59,413 | 59,413 | 0,000 | 0,0859 | 2,7491 | | | 59,41315 | |
| 15 | 3,331 | 58,619 | 58,619 | 58,619 | 0,000 | 0,0848 | 3,1362 | | | 58,61902 | |
| 15 | 4,631 | 57,318 | 57,318 | 57,318 | 0,000 | 0,0829 | 3,4810 | | | 57,31837 | |
| 15 | 7,247 | 54,703 | 54,703 | 54,703 | 0,000 | 0,0791 | 3,7176 | | | 54,70267 | |
| 15 | 11,721 | 50,229 | 50,229 | 50,229 | 0,000 | 0,0726 | 3,7767 | | | 50,22874 | |
| 15 | 20,107 | 41,843 | 41,843 | 41,843 | 0,000 | 0,0605 | 3,4487 | | | 41,84294 | |
| 15 | 27,446 | 34,503 | 34,503 | 34,503 | 0,000 | 0,0499 | 3,0932 | | | 34,50324 | |
| 15 | 38,218 | 23,732 | 23,732 | 23,732 | 0,000 | 0,0343 | 2,2992 | | | 23,73194 | |
| 0 | 0,000 | 0,000 | 0,000 | 0,000 | 0,000 | 0,0000 | 0,0000 | | | 0,00000 | |
| 0 | 0,000 | 0,000 | 0,000 | 2,315 | 2,315 | 0,0000 | 0,0000 | | | 0,00000 | |
| 0 | 0,000 | 0,000 | 0,000 | 30,663 | 30,663 | 0,0000 | 0,0000 | | | 0,00000 | |
| 0 | 0,000 | 0,000 | 0,000 | 60,995 | 60,995 | 0,0000 | 0,0000 | | | 0,00000 | |
| 0 | 0,000 | 0,000 | 0,000 | 99,395 | 99,395 | 0,0000 | 0,0000 | | | 0,00000 | |
| 0 | 0,000 | 0,000 | 0,000 | 137,149 | 137,149 | 0,0000 | 0,0000 | | | 0,00000 | |
| 0 | 0,000 | 0,000 | 0,000 | 172,183 | 172,183 | 0,0000 | 0,0000 | | | 0,00000 | |
| 0 | 0,000 | 0,000 | 3,067 | 3,067 | 0,000 | 0,0192 | 0,0192 | 35,67 | 474 | 3,06662 | 779,8 |
| 0 | 0,000 | 0,000 | 19,066 | 19,066 | 0,000 | 0,0299 | 0,0897 | | | 19,06632 | |
| 0 | 0,000 | 0,000 | 40,808 | 40,808 | 0,000 | 0,0512 | 0,3585 | | | 40,80762 | |
| 0 | 0,000 | 0,000 | 53,345 | 53,345 | 0,000 | 0,0669 | 0,8033 | | | 53,34467 | |
| 0 | 0,000 | 0,000 | 58,397 | 58,397 | 0,000 | 0,0733 | 1,2458 | | | 58,39679 | |
| 0 | 0,000 | 0,000 | 60,658 | 60,658 | 0,000 | 0,0761 | 1,6747 | | | 60,65816 | |
| 0 | 0,000 | 0,000 | 63,462 | 63,462 | 0,000 | 0,0796 | 2,1503 | | | 63,46189 | |

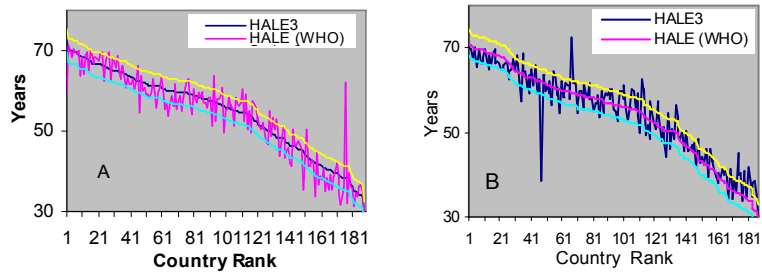

Fig. 6. Application for WHO data, males 2000
A. Ranking according to $HALE_3$,
B. Ranking according to HALE (WHO)

The $HALE_3$ results of the application to the WHO data for 2000 for males are summarized in Table III. In the same Table we have included the HALE (WHO) data provided from the WHO report. The $HALE_3$ where preferred for comparison as are more close to the HALE (WHO) estimates. Figure 6A illustrates the country ranking according to $HALE_3$ and Figure 6B the country ranking according to HALE (WHO). In both cases the estimates are close enough and we have checked a 3 year confidence interval quite reasonable if we take into consideration the uncertainty in the estimates presented in the HALE (WHO) report. Grenada and Quatar estimates are far from the confidence intervals while the other countries' behavior supports our methodology. Figure 7 illustrates the $HALE_3$ and HALE (WHO) when the country ranking is according to the life expectancy at birth $e_x$.

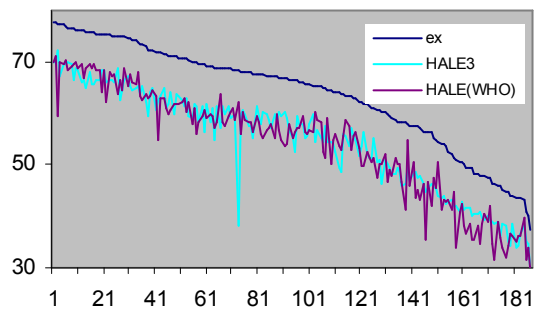

Fig. 7. WHO data, males 2000, ranking according to $e_x$

Table III

| Rank | Country Males 2000 | HALE3 | HALE (WHO) | Rank | Country Males 2000 | HALE3 | HALE (WHO) | Rank | Country Males 2000 | HALE3 | HALE (WHO) |
|---|---|---|---|---|---|---|---|---|---|---|---|
| 1 | Japan | 69,9 | 71,2 | 65 | Romania | 61,3 | 59,5 | 129 | Sao Tome and P | 54,4 | 50,3 |
| 2 | Switzerland | 69,6 | 70,4 | 66 | Suriname | 58,0 | 59,5 | 130 | Russian Federat | 49,2 | 50,3 |
| 3 | Sweden | 67,4 | 70,1 | 67 | Qatar | 72,4 | 59,3 | 131 | Mongolia | 46,2 | 50,3 |
| 4 | Iceland | 69,9 | 69,8 | 68 | Poland | 61,4 | 59,3 | 132 | Pakistan | 50,8 | 50,2 |
| 5 | Andorra | 68,7 | 69,8 | 69 | Oman | 61,8 | 59,2 | 133 | Bhutan | 48,5 | 50,1 |
| 6 | Greece | 65,5 | 69,7 | 70 | Iran (Islamic Rep | 54,4 | 59,0 | 134 | Kyrgyzstan | 56,3 | 49,6 |
| 7 | Australia | 69,6 | 69,6 | 71 | Fiji | 56,7 | 58,7 | 135 | Tajikistan | 53,5 | 49,6 |
| 8 | Italy | 66,5 | 69,5 | 72 | Mauritius | 60,0 | 58,6 | 136 | Yemen | 49,5 | 48,9 |
| 9 | New Zealand | 66,1 | 69,5 | 73 | Colombia | 59,6 | 58,6 | 137 | Myanmar | 48,4 | 47,7 |
| 10 | Monaco | 68,0 | 69,4 | 74 | Sri Lanka | 55,7 | 58,6 | 138 | Nepal | 50,4 | 47,5 |
| 11 | Israel | 69,5 | 69,3 | 75 | Libyan Arab Jam | 61,4 | 58,4 | 139 | Gambia | 45,4 | 47,3 |
| 12 | Denmark | 66,6 | 68,9 | 76 | Algeria | 59,4 | 58,4 | 140 | Gabon | 45,8 | 46,8 |
| 13 | Norway | 64,9 | 68,8 | 77 | Ecuador | 58,2 | 58,4 | 141 | Papua New Guin | 50,2 | 46,6 |
| 14 | Malta | 68,3 | 68,7 | 78 | Saudi Arabia | 61,5 | 58,3 | 142 | Ghana | 45,4 | 46,5 |
| 15 | Spain | 65,7 | 68,7 | 79 | Viet Nam | 57,7 | 58,2 | 143 | Comoros | 46,2 | 46,2 |
| 16 | France | 68,5 | 68,5 | 80 | Jordan | 57,6 | 58,2 | 144 | Sudan | 47,2 | 45,7 |
| 17 | Canada | 69,8 | 68,3 | 81 | Samoa | 55,0 | 58,2 | 145 | Cambodia | 45,4 | 45,6 |
| 18 | United Kingdom | 66,5 | 68,3 | 82 | Solomon Islands | 57,1 | 58,0 | 146 | Senegal | 47,3 | 45,2 |
| 19 | Netherlands | 66,6 | 68,2 | 83 | Belize | 56,9 | 58,0 | 147 | Equatorial Guine | 41,9 | 44,9 |
| 20 | Austria | 66,5 | 68,1 | 84 | Peru | 58,9 | 57,8 | 148 | Lao People's De | 48,4 | 43,7 |
| 21 | Ireland | 65,1 | 67,8 | 85 | Thailand | 53,5 | 57,7 | 149 | Madagascar | 47,6 | 43,2 |
| 22 | Belgium | 65,1 | 67,7 | 86 | Saint Kitts and N | 59,7 | 57,6 | 150 | Benin | 42,7 | 43,1 |
| 23 | Luxembourg | 66,0 | 67,6 | 87 | Bahamas | 54,6 | 57,2 | 151 | South Africa | 44,1 | 43,0 |
| 24 | Germany | 66,5 | 67,4 | 88 | Egypt | 58,0 | 57,1 | 152 | Togo | 42,5 | 42,7 |
| 25 | Singapore | 66,6 | 66,8 | 89 | Philippines | 59,4 | 57,0 | 153 | Congo | 42,6 | 42,5 |
| 26 | Cyprus | 62,6 | 66,4 | 90 | Seychelles | 58,7 | 57,0 | 154 | Mauritania | 46,3 | 42,1 |
| 27 | Finland | 64,5 | 66,1 | 91 | Armenia | 60,2 | 56,9 | 155 | Nigeria | 39,4 | 42,1 |
| 28 | United States of | 64,9 | 65,7 | 92 | Cape Verde | 58,0 | 56,9 | 156 | Eritrea | 46,2 | 41,4 |
| 29 | Cuba | 67,4 | 65,1 | 93 | Turkey | 60,0 | 56,8 | 157 | Haiti | 43,4 | 41,3 |
| 30 | Kuwait | 64,6 | 64,6 | 94 | Indonesia | 55,4 | 56,5 | 158 | Kenya | 43,0 | 41,2 |
| 31 | Slovenia | 64,7 | 64,5 | 95 | Albania | 55,0 | 56,5 | 159 | Cameroon | 41,0 | 40,9 |
| 32 | Costa Rica | 68,4 | 64,2 | 96 | Palau | 52,7 | 56,5 | 160 | Guinea | 39,5 | 40,4 |
| 33 | Portugal | 62,1 | 63,9 | 97 | Tuvalu | 54,7 | 56,4 | 161 | Zimbabwe | 38,2 | 39,6 |
| 34 | The Former Yugo | 57,3 | 63,9 | 98 | Estonia | 59,7 | 56,2 | 162 | Côte d'Ivoire | 40,2 | 39,1 |
| 35 | Brunei Darussala | 68,6 | 63,8 | 99 | Georgia | 59,8 | 56,1 | 163 | Swaziland | 39,1 | 38,8 |
| 36 | Chile | 63,7 | 63,5 | 100 | Vanuatu | 57,8 | 56,0 | 164 | United Republic | 41,5 | 38,6 |
| 37 | Republic of Kore | 64,3 | 63,2 | 101 | Nicaragua | 61,2 | 55,8 | 165 | Chad | 39,7 | 38,6 |
| 38 | Dominica | 60,8 | 63,2 | 102 | Micronesia (Fed | 55,6 | 55,8 | 166 | Liberia | 40,5 | 38,2 |
| 39 | Mexico | 63,3 | 63,1 | 103 | Honduras | 52,4 | 55,8 | 167 | Botswana | 41,3 | 38,1 |
| 40 | Bahrain | 59,4 | 63,0 | 104 | Republic of Mold | 55,8 | 55,4 | 168 | Guinea-Bissau | 34,1 | 36,7 |
| 41 | Czech Republic | 64,7 | 62,9 | 105 | Belarus | 55,7 | 55,4 | 169 | Namibia | 41,5 | 36,5 |
| 42 | Jamaica | 61,4 | 62,9 | 106 | Hungary | 60,4 | 55,3 | 170 | Uganda | 36,7 | 36,2 |
| 43 | Panama | 65,4 | 62,6 | 107 | Morocco | 59,7 | 55,3 | 171 | Angola | 34,3 | 36,2 |
| 44 | United Arab Emir | 66,1 | 62,3 | 108 | El Salvador | 55,3 | 55,3 | 172 | Lesotho | 38,4 | 36,1 |
| 45 | Barbados | 59,4 | 62,3 | 109 | Brazil | 59,5 | 54,9 | 173 | Ethiopia | 36,3 | 35,7 |
| 46 | Bosnia and Herz | 62,7 | 62,1 | 110 | Dem. Peoples's | 54,1 | 54,9 | 174 | Djibouti | 44,9 | 35,6 |
| 47 | Grenada | 38,3 | 62,1 | 111 | Marshall Islands | 49,9 | 54,8 | 175 | Somalia | 40,0 | 35,5 |
| 48 | Argentina | 63,4 | 61,8 | 112 | Dominican Repu | 62,5 | 54,7 | 176 | Burkina Faso | 40,3 | 35,4 |
| 49 | Uruguay | 63,7 | 61,7 | 113 | Maldives | 59,1 | 54,2 | 177 | Afghanistan | 33,8 | 35,1 |
| 50 | Antigua and Bart | 59,7 | 61,7 | 114 | Lithuania | 60,1 | 53,6 | 178 | Mali | 40,8 | 34,8 |
| 51 | Tunisia | 61,2 | 61,0 | 115 | Guatemala | 49,3 | 53,5 | 179 | Central African F | 39,0 | 34,7 |
| 52 | Bulgaria | 59,4 | 61,0 | 116 | Azerbaijan | 51,6 | 53,3 | 180 | Dem. Rep. of the | 34,9 | 34,4 |
| 53 | China | 61,9 | 60,9 | 117 | Kiribati | 48,5 | 52,8 | 181 | Niger | 41,9 | 33,9 |
| 54 | Croatia | 56,3 | 60,8 | 118 | Uzbekistan | 57,2 | 52,7 | 182 | Burundi | 38,5 | 33,9 |
| 55 | Saint Lucia | 64,3 | 60,7 | 119 | Iraq | 54,5 | 52,6 | 183 | Zambia | 34,7 | 33,7 |
| 56 | Venezuela | 62,8 | 60,4 | 120 | Ukraine | 52,3 | 52,3 | 184 | Rwanda | 38,2 | 32,0 |
| 57 | Cook Islands | 58,9 | 60,4 | 121 | India | 50,8 | 52,2 | 185 | Mozambique | 38,7 | 31,5 |
| 58 | Lebanon | 60,5 | 60,3 | 122 | Latvia | 57,4 | 51,4 | 186 | Malawi | 35,0 | 31,4 |
| 59 | Trinidad and Tob | 54,5 | 60,3 | 123 | Guyana | 55,1 | 51,4 | 187 | Sierra Leone | 29,7 | 29,7 |
| 60 | Paraguay | 62,0 | 59,9 | 124 | Bolivia | 51,2 | 51,4 | | | | |
| 61 | Saint Vincent and | 58,2 | 59,7 | 125 | Turkmenistan | 48,1 | 51,2 | | | | |
| 62 | Malaysia | 58,1 | 59,7 | 126 | Bangladesh | 51,2 | 50,6 | | | | |
| 63 | Syrian Arab Rep | 61,4 | 59,6 | 127 | Kazakhstan | 48,1 | 50,5 | | | | |
| 64 | Slovakia | 61,2 | 59,6 | 128 | Nauru | 43,9 | 50,4 | | | | |

Another very interesting point is to check our HALE estimates to the estimates published in *The Lancet* by *Colin D Mathers, Ritu Sadana, Joshua A Salomon, Christopher JL Murray, Alan D Lopez,* 2001 in a paper titled *Healthy life expectancy in 191 countries, 1999*.

Mathers et al. study is based on the 1999 WHO data. For our comparison we have the 2000 data from WHO. However, the one year interval is quite small and we proceed in comparing both results presented in Table IV. We keep the notation of the authors for the healthy life expectancy as DALE in the Table IV and in the following Figure 8 where the WHO member countries are ranked according to the life expectancy at birth. Our estimates from $HALE_3$ estimator are in relative accordance to DALE estimated by Mathers et al.

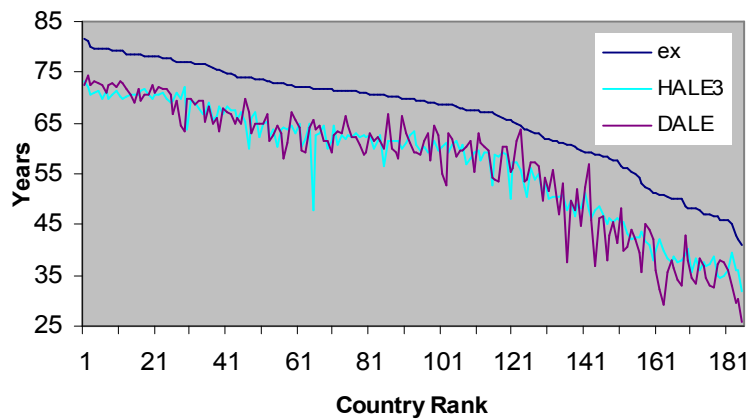

Fig. 8. DALE 1999, $HALE_3$ 2000 and $e_x$ 2000

**Conclusions**

We have proposed and applied a method to estimate the Health State Function of a Population and various related characteristics and estimators. We have used the related theoretical framework to estimate the Healthy Life Expectancy of the population and we have done applications by using the life table data for the member countries of the World Health Organization by comparing our results with the provided from WHO. The findings indicate that the proposed methodology can be a useful tool for estimating the "loss of healthy life years" and then to find the Healthy Life Expectancy of a population. Even more, by these applications, we straighten the proposed theory for the health state of a population. We do believe that interesting applications is various fields may arise.

Table IV

| Rank | Country | HALE3 2000 | DALE 1999 | Rank | Country | HALE3 2000 | DALE 1999 | Rank | Country | HALE3 2000 | DALE 1999 |
|---|---|---|---|---|---|---|---|---|---|---|---|
| 1 | Japan | 71,9 | 74,5 | 65 | Cook Islands | 60,8 | 63,4 | 129 | Sao Tome and Pr | 58,2 | 53,5 |
| 2 | Australia | 70,8 | 73,2 | 66 | Kuwait | 68,4 | 63,2 | 130 | Bolivia | 52,2 | 53,3 |
| 3 | France | 70,4 | 73,1 | 67 | Estonia | 62,5 | 63,1 | 131 | India | 50,9 | 53,2 |
| 4 | Sweden | 71,4 | 73,0 | 68 | Paraguay | 66,0 | 63,0 | 132 | Vanuatu | 60,9 | 52,8 |
| 5 | Spain | 69,7 | 72,8 | 69 | Oman | 62,5 | 63,0 | 133 | Nauru | 51,3 | 52,5 |
| 6 | Italy | 70,6 | 72,7 | 70 | Ukraine | 59,5 | 63,0 | 134 | Bhutan | 49,0 | 51,8 |
| 7 | Switzerland | 70,5 | 72,5 | 71 | Colombia | 63,8 | 62,9 | 135 | Myanmar | 50,1 | 51,6 |
| 8 | Greece | 69,6 | 72,5 | 72 | Turkey | 61,0 | 62,9 | 136 | Bangladesh | 50,4 | 49,9 |
| 9 | Monaco | 69,9 | 72,4 | 73 | Tonga | 59,2 | 62,9 | 137 | Yemen | 49,6 | 49,7 |
| 10 | San Marino | 73,1 | 72,3 | 74 | Sri Lanka | 59,1 | 62,8 | 138 | Nepal | 50,6 | 49,5 |
| 11 | Andorra | 69,6 | 72,3 | 75 | Mauritius | 64,5 | 62,7 | 139 | Gambia | 45,4 | 48,3 |
| 12 | Canada | 71,3 | 72,0 | 76 | Suriname | 59,5 | 62,7 | 140 | Gabon | 46,5 | 47,8 |
| 13 | Netherlands | 70,7 | 72,0 | 77 | Dominican Repu | 63,7 | 62,5 | 141 | Papua New Guine | 50,4 | 47,0 |
| 14 | United Kingdor | 70,8 | 71,7 | 78 | Romania | 63,0 | 62,3 | 142 | Comoros | 47,2 | 46,8 |
| 15 | Norway | 70,2 | 71,7 | 79 | China | 62,2 | 62,3 | 143 | Lao People's Den | 48,5 | 46,1 |
| 16 | Austria | 70,4 | 71,6 | 80 | Latvia | 61,6 | 62,2 | 144 | Cambodia | 46,2 | 45,7 |
| 17 | Belgium | 69,7 | 71,6 | 81 | Belarus | 60,7 | 61,7 | 145 | Ghana | 45,6 | 45,5 |
| 18 | Luxembourg | 70,6 | 71,1 | 82 | Niue | 64,4 | 61,6 | 146 | Congo | 41,6 | 45,1 |
| 19 | Iceland | 71,3 | 70,8 | 83 | Algeria | 60,3 | 61,6 | 147 | Senegal | 48,0 | 44,6 |
| 20 | Malta | 71,7 | 70,5 | 84 | Saint Kitts and N | 60,0 | 61,6 | 148 | Equatorial Guinea | 40,9 | 44,1 |
| 21 | Finland | 69,1 | 70,5 | 85 | El Salvador | 62,6 | 61,5 | 149 | Haiti | 42,1 | 43,8 |
| 22 | Germany | 70,6 | 70,4 | 86 | Republic of Mol | 58,1 | 61,5 | 150 | Sudan | 46,4 | 43,0 |
| 23 | Israel | 70,5 | 70,4 | 87 | Tunisia | 62,3 | 61,4 | 151 | Côte d'Ivoire | 38,9 | 42,8 |
| 24 | United States ( | 69,3 | 70,0 | 88 | Malaysia | 59,9 | 61,4 | 152 | Benin | 42,6 | 42,2 |
| 25 | Dominica | 65,3 | 69,8 | 89 | Russian Federa | 57,2 | 61,3 | 153 | Cameroon | 37,9 | 42,2 |
| 26 | Cyprus | 63,6 | 69,8 | 90 | Honduras | 57,8 | 61,1 | 154 | Mauritania | 46,3 | 41,4 |
| 27 | Ireland | 66,6 | 69,6 | 91 | Ecuador | 63,8 | 61,0 | 155 | Togo | 43,1 | 40,7 |
| 28 | Denmark | 70,8 | 69,4 | 92 | Belize | 62,7 | 60,9 | 156 | South Africa | 45,7 | 39,8 |
| 29 | Singapore | 71,4 | 69,3 | 93 | Lebanon | 62,0 | 60,6 | 157 | Kenya | 42,4 | 39,3 |
| 30 | Portugal | 68,1 | 69,3 | 94 | Iran (Islamic Re | 59,3 | 60,5 | 158 | Nigeria | 35,9 | 38,3 |
| 31 | New Zealand | 70,7 | 69,2 | 95 | Uzbekistan | 59,0 | 60,2 | 159 | Swaziland | 40,1 | 38,1 |
| 32 | Chile | 69,0 | 68,6 | 96 | Guyana | 58,6 | 60,2 | 160 | Angola | 34,7 | 38,0 |
| 33 | Slovenia | 69,2 | 68,4 | 97 | Thailand | 57,0 | 60,2 | 161 | Djibouti | 45,2 | 37,9 |
| 34 | Czech Republi | 66,7 | 68,0 | 98 | Jordan | 61,5 | 60,0 | 162 | Guinea | 37,6 | 37,8 |
| 35 | Jamaica | 64,3 | 67,3 | 99 | Albania | 56,4 | 60,0 | 163 | Eritrea | 47,8 | 37,7 |
| 36 | Uruguay | 68,2 | 67,0 | 100 | Indonesia | 59,4 | 59,7 | 164 | Afghanistan | 34,7 | 37,7 |
| 37 | Croatia | 59,8 | 67,0 | 101 | Micronesia (Fed | 59,0 | 59,6 | 165 | Guinea-Bissau | 35,0 | 37,2 |
| 38 | Costa Rica | 69,5 | 66,7 | 102 | Fiji | 60,8 | 59,4 | 166 | Lesotho | 37,6 | 36,9 |
| 39 | Argentina | 67,7 | 66,7 | 103 | Peru | 60,4 | 59,4 | 167 | Madagascar | 47,9 | 36,6 |
| 40 | Armenia | 61,6 | 66,7 | 104 | Libyan Arab Jan | 62,4 | 59,3 | 168 | Somalia | 38,8 | 36,4 |
| 41 | Slovakia | 65,4 | 66,6 | 105 | Seychelles | 61,5 | 59,3 | 169 | United Republic c | 40,0 | 36,0 |
| 42 | Saint Vincent a | 59,8 | 66,4 | 106 | Bahamas | 61,9 | 59,1 | 170 | Central African R | 35,9 | 36,0 |
| 43 | Georgia | 61,7 | 66,3 | 107 | Brazil | 61,3 | 59,1 | 171 | Namibia | 43,6 | 35,6 |
| 44 | Poland | 65,5 | 66,2 | 108 | Morocco | 60,8 | 59,1 | 172 | Burkina Faso | 38,3 | 35,5 |
| 45 | Panama | 66,5 | 66,0 | 109 | Palau | 63,2 | 59,0 | 173 | Burundi | 36,8 | 34,6 |
| 46 | Antigua and Ba | 63,1 | 65,8 | 110 | Philippines | 59,7 | 58,9 | 174 | Mozambique | 35,7 | 34,4 |
| 47 | Grenada | 47,7 | 65,5 | 111 | Syrian Arab Rep | 62,5 | 58,8 | 175 | Liberia | 37,5 | 34,0 |
| 48 | United Arab Er | 66,7 | 65,4 | 112 | Egypt | 61,3 | 58,5 | 176 | Ethiopia | 38,3 | 33,5 |
| 49 | Mexico | 67,5 | 65,0 | 113 | Viet Nam | 61,5 | 58,2 | 177 | Mali | 37,9 | 33,1 |
| 50 | Saint Lucia | 67,2 | 65,0 | 114 | Nicaragua | 64,1 | 58,1 | 178 | Zimbabwe | 39,3 | 32,9 |
| 51 | Republic of Ko | 65,8 | 65,0 | 115 | Cape Verde | 59,4 | 57,6 | 179 | Rwanda | 37,2 | 32,8 |
| 52 | Venezuela | 64,7 | 65,0 | 116 | Tuvalu | 53,8 | 57,4 | 180 | Uganda | 38,7 | 32,7 |
| 53 | Barbados | 62,1 | 65,0 | 117 | Tajikistan | 56,1 | 57,3 | 181 | Botswana | 42,1 | 32,3 |
| 54 | Bosnia and He | 67,1 | 64,9 | 118 | Marshall Islands | 47,7 | 56,8 | 182 | Zambia | 36,0 | 30,3 |
| 55 | Trinidad and T | 61,3 | 64,6 | 119 | Kazakhstan | 54,8 | 56,4 | 183 | Malawi | 36,0 | 29,4 |
| 56 | Saudi Arabia | 62,5 | 64,5 | 120 | Kyrgyzstan | 57,4 | 56,3 | 184 | Niger | 39,8 | 29,1 |
| 57 | Brunei Darussa | 69,7 | 64,4 | 121 | Pakistan | 50,3 | 55,9 | 185 | Sierra Leone | 31,8 | 25,9 |
| 58 | Bulgaria | 63,0 | 64,4 | 122 | Iraq | 59,2 | 55,3 | | | | |
| 59 | Bahrain | 60,3 | 64,4 | 123 | Kiribati | 50,1 | 55,3 | | | | |
| 60 | Lithuania | 65,0 | 64,1 | 124 | Solomon Islands | 60,5 | 54,9 | | | | |
| 61 | Hungary | 64,8 | 64,1 | 125 | Turkmenistan | 53,7 | 54,3 | | | | |
| 62 | The Former Yu | 62,7 | 63,7 | 126 | Guatemala | 52,8 | 54,3 | | | | |
| 63 | Azerbaijan | 55,8 | 63,7 | 127 | Maldives | 58,9 | 53,9 | | | | |
| 64 | Qatar | 72,2 | 63,5 | 128 | Mongolia | 50,3 | 53,8 | | | | |